\documentclass[aps,twocolumn,pra,superscriptaddress,letterpaper,nofootinbib,longbibliography]{revtex4-2}
\usepackage{amssymb}
\usepackage{physics}
\usepackage{mathtools}
\usepackage{upgreek}
\usepackage{booktabs}
\usepackage{siunitx}
\usepackage{url}
\AtBeginDocument{
\heavyrulewidth=.08em
\lightrulewidth=.05em
\cmidrulewidth=.03em
\belowrulesep=.65ex
\belowbottomsep=0pt
\aboverulesep=.4ex
\abovetopsep=0pt
\cmidrulesep=\doublerulesep
\cmidrulekern=.5em
\defaultaddspace=.5em
}

\usepackage{txfonts}
\usepackage[pdftex]{graphicx}
\usepackage[usenames, dvipsnames, svgnames, table]{xcolor}
\usepackage[colorlinks=true, citecolor=Blue,linkcolor=BrickRed, urlcolor=magenta]{hyperref}
\usepackage{bm}
\usepackage{placeins} 
\usepackage{setspace}
\usepackage[final]{changes}

\usepackage{cleveref}
\crefname{section}{Section}{Sections}
\crefname{figure}{Figure}{Figures}
\crefname{table}{Table}{Tables}
\crefname{equation}{Eq.}{Eq.}

\newcommand{\e}{{\rm e}}

\newcommand{\eps}{\epsilon}                                                
\newcommand{\rmsb}{{\rm sb}}                                                
\newcommand{\mix}{{\rm mix}}                                                

\newcommand{\T}[1]{\mathbf{T}_{#1}}
\newcommand{\Tinv}[1]{\mathbf{T}^{-1}_{#1}}

\renewcommand{\S}{\mathbf{S}}

\newcommand{\F}{\mathbf{F}}

\newcommand{\qt}[1]{q_{#1}}                                     		   

\newcommand{\qdt}[1]{\dot{q}_{#1}}                                     	   

\newcommand{\rt}[3]{r^{#1}_{{#2}{#3}}}             

\newcommand{\rdt}[3]{\dot{r}^{#1}_{{#2}{#3}}}             

\begin{document}
\title{Experimental verification of intersatellite clock synchronization \\ at LISA performance levels}

\author{Kohei Yamamoto}
\email{kohei.yamamoto@aei.mpg.de}
\affiliation{Max-Planck Institut f\"ur Gravitationsphysik (Albert-Einstein Institut),\\
Callinstra\ss e 38, 30167 Hannover, Germany}

\author{Christoph Vorndamme}
\affiliation{Max-Planck Institut f\"ur Gravitationsphysik (Albert-Einstein Institut),\\
Callinstra\ss e 38, 30167 Hannover, Germany}

\author{Olaf Hartwig}
\affiliation{Max-Planck Institut f\"ur Gravitationsphysik (Albert-Einstein Institut),\\
Callinstra\ss e 38, 30167 Hannover, Germany}
\affiliation{LNE-SYRTE, Observatoire de Paris, Université PSL, CNRS, Sorbonne Université,
61 avenue de l’Observatoire, 75014 Paris, France}

\author{Martin Staab}
\affiliation{Max-Planck Institut f\"ur Gravitationsphysik (Albert-Einstein Institut),\\
Callinstra\ss e 38, 30167 Hannover, Germany}

\author{Thomas S. Schwarze}
\email{thomas.schwarze@aei.mpg.de}
\affiliation{Max-Planck Institut f\"ur Gravitationsphysik (Albert-Einstein Institut),\\
Callinstra\ss e 38, 30167 Hannover, Germany}

\author{Gerhard Heinzel}
\affiliation{Max-Planck Institut f\"ur Gravitationsphysik (Albert-Einstein Institut),\\
Callinstra\ss e 38, 30167 Hannover, Germany}

\begin{abstract}
The Laser Interferometer Space Antenna (LISA) aims to observe gravitational waves in the \si{\milli\Hz} regime over its 10-year mission time. LISA will operate laser interferometers between three spacecrafts. Each spacecraft will utilize independent clocks which determine the sampling times of onboard phasemeters to extract the interferometric phases and, ultimately, gravitational wave signals. To suppress limiting laser frequency noise, signals sampled by each phasemeter need to be combined in postprocessing to synthesize virtual equal-arm interferometers. The synthesis in turn requires a synchronization of the independent clocks. This article reports on the experimental verification of a clock synchronization scheme down to LISA performance levels using a hexagonal optical bench. The development of the scheme includes data processing that is expected to be applicable to the real LISA data with minor modifications. Additionally, some noise coupling mechanisms are discussed.
\end{abstract}

\maketitle

\section{Introduction}
The first detection of the gravitational waves (GWs) by the Laser Interferometer Gravitational-Wave Observatory (LIGO) and Virgo in 2015 was the dawn of the gravitational wave astronomy \cite{GW150914}. The target observation band of these ground-based detectors is \SI{1}{\Hz} to \SI{1}{\kilo\Hz}, being limited by seismic and gravity gradient noise below \SI{1}{\Hz}.

The Laser Interferometer Space Antenna (LISA), being a gravitational-wave detector in space, will avoid the mentioned limitations, targeting the observation band from \SI{0.1}{\milli\Hz} to \SI{1}{\Hz}. This mission is composed of three spacecraft (SC), forming a triangle with 2.5 million \si{\kilo\meter} arm lengths. Each SC hosts a free-falling test mass (TM). The microscopic relative displacement of these TMs will be sensed using intersatellite heterodyne laser interferometry with \SI{10}{\pico\meter\per\sqrt{\rm\Hz}} precision per TM pair. Hence, GW signals will be detectable in the interferometric phases extracted by digital phasemeters on each SC. The sample timing of the latter in turn is determined by onboard clocks, namely one per SC and all independent from each other.

Orbital variations will lead to arm length drifts of \SI{10}{\meter\per\second} as well as arm length mismatches by the order of $10^{8}$\,\si{\meter}. Such unequal arm lengths cause a large coupling of laser frequency noise into the interferometric phase readout. To mitigate this overwhelming noise source, a virtual interferometer insensitive to laser frequency noise can be synthesised in postprocessing by a technique named time-delay interferometry (TDI) \cite{Tinto1999} . This scheme requires to shift phasemeter (PM) signals by precise time intervals related to the light travel times along the arms, ideally in a common clock frame. This is impeded by the uncertain relation of sample times between different SC due to the independent clocks, which exhibit offsets and drifts, and the light travel delays between the SC. On top, differential clock jitter in the \si{\milli\Hz} measurement band directly couples into phase sensing precision. 

State-of-the-art space-qualified oscillators have a typical Allan deviation of about $10^{-13}$ for averaging times of \SI{1}{\second} \cite{Weaver2010}. That is around 3-4 orders larger than the required level to overcome the differential clock jitter \cite{Tinto2001,Hartwig2021}. Hence, LISA requires a scheme to deal with both differential clock jitters and synchronizing sample timing of all signals in postprocessing. The total process is summarized as clock synchronization in the following. It leads to the necessity of auxiliary monitors of the differential clocks in addition to the main measurements of optical carrier-carrier beatnotes \cite{PrePhaseA, Heinzel2011}. These are implemented by encoding the clock tones in optical sidebands of the beams. The optical beatnote between the local and received sideband contains the differential clock information. To relax phase fidelity requirements on the clock tone transfer, the tones are  up-converted from the \si{\mega\Hz} to the \si{\giga\Hz} regime before being imprinted on the heterodyne beatnote as phase modulation. Furthermore, to complete the clock synchronization, the aforementioned delays need to be derived. On top of the baseline solution based on pseudo-random-noise (PRN) ranging \cite{Heinzel2011, Esteban2011}, TDI ranging (TDIR) \cite{Tinto2005} can act as a complementary scheme. PRN ranging encodes a pseudo-random noise via phase modulations in addition to the aforementioned clock sideband. SC separations and the differences between distant and local clock times are derived by comparing the received PRN code and a local copy of it. In contrast, TDIR derives these parameters by minimizing the noise level after TDI processing. 

This article reports on the experimental demonstration of the clock synchronization among three independent clocks. To demonstrate LISA performance levels (\SI{1}{\pico\meter\per\sqrt{\rm\Hz}} per single PM readout channel with a noise shape function \cite{LISA_L3}), an implementation of a three-signal scheme by a picometer-stable hexagonal optical bench \cite{Schwarze2019} is used. It demonstrates the suppression of noise due to unsynchronized clocks by around 6 orders of magnitude, enabling the performance of the optical three-signal combination on LISA levels. While the setup cannot feature realistic light travel delays, the absence of a synchronization start pulse \cite{Vine2010} between PMs inserts a realistic unknown initial delay. Consequently, these delays have been determined by TDIR-like processing. Note that PRN ranging is not implemented in this experiment yet.

Besides the experimental results, a detailed model of the clock synchronization based on the measured total frequencies is given, which is expected to be applicable with minor modifications also to the real LISA data \cite{HartwigUnpublished}. Moreover, noise couplings stemming from the clock synchronization are discussed.


\section{Experimental setup}
\label{sec:setup}
\begin{figure*}
\begin{centering}
\includegraphics[width=12.9cm]{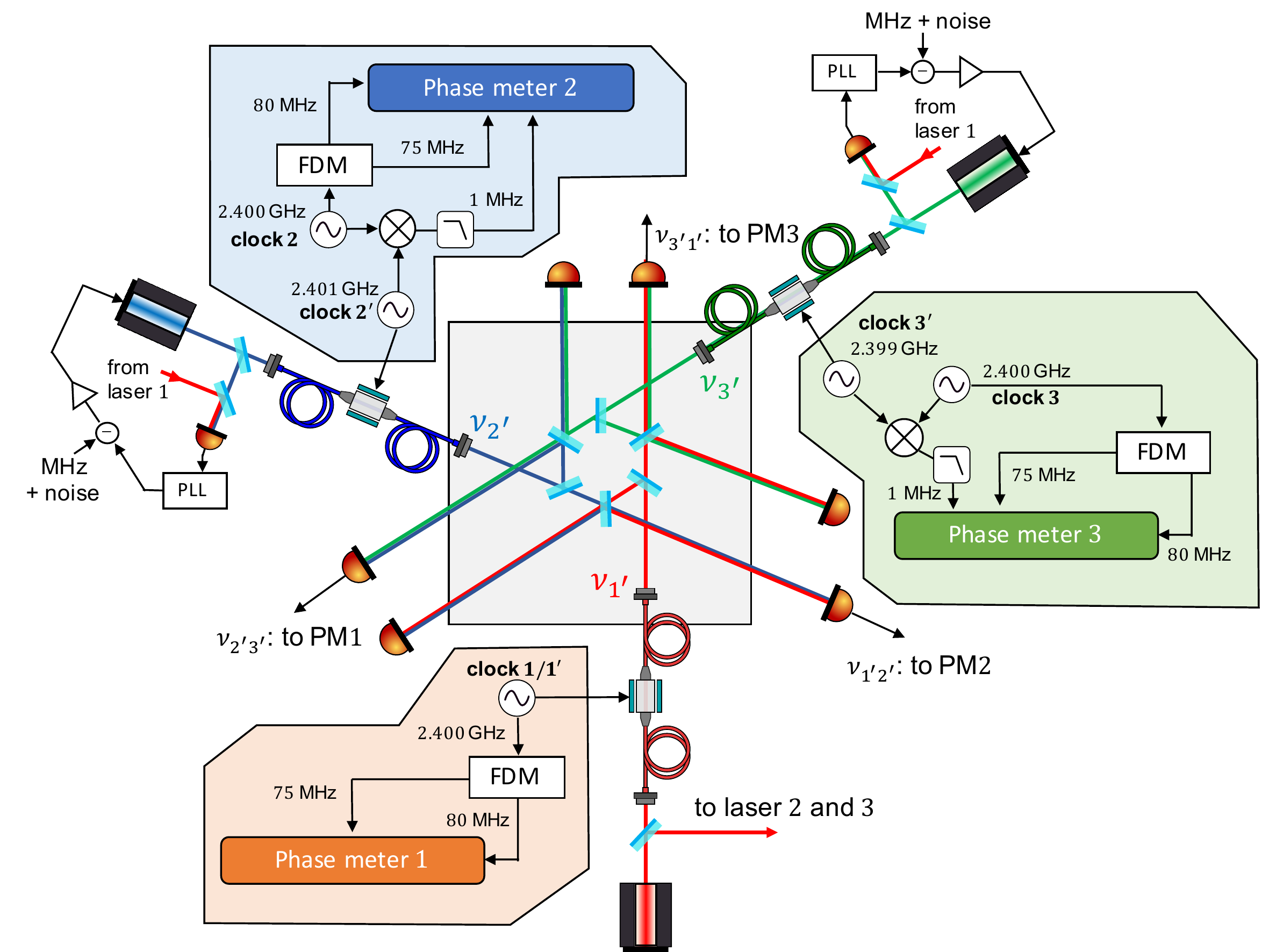}
\caption{\label{fig:layout} Schematic of the experimental setup to demonstrate the clock synchronization at LISA performance levels. The three lasers are locked to each other with \si{\mega\Hz} offsets and noise injections interfere pairwise and generate three optical beatnotes. Their combination should cancel out by design and allows the characterization of the metrology chain, e.g. including the three readout PMs. A total of five clocks as well as three optical and two electrical beatnotes are used to achieve synchronization between PMs.}
\end{centering}
\end{figure*}

The experimental setup applying the clock synchronization in a hexagonal optical testbed was first proposed in \cite{Schwarze2016} and is illustrated in \cref{fig:layout}. Conceptually, the three colored phasemeter-clock assemblies together with the optical devices like lasers and electro-optic modulators (EOM) can be interpreted as the three LISA SC. 

Three optical heterodyne beatnotes are generated pairwise from the three lasers. Two of these act as secondaries locked to the remaining primary with LISA-like \si{\mega\Hz} offsets. In this article, the \si{\mega\Hz} offsets are time-invariant. The beatnotes are consecutively detected by pairs of complementary photoreceivers and fed to PMs. By design, the phases of the three beatnotes should cancel out in a three-signal combination \cite{Shaddock2006, Schwarze2019}. Residual noise gives a measure of the PM and metrology chain performance.

The setup features five independent clocks in total, all of which have frequencies around \SI{2.4}{\giga\Hz}. They are categorized into two groups: unprimed clocks for directly driving PMs and primed clocks for driving EOMs. The latter are responsible for creating \si{\giga\Hz} sidebands via phase modulations. 

Each of the unprimed clocks, running at \SI{2.400}{\giga\Hz}, drives a phasemeter module called frequency distribution module (FDM) \cite{FinalReport}. Each FDM in turn consists of two frequency-divider chains: one to derive \SI{75}{\mega\Hz} pilot-tone signals (PT) \cite{Shaddock2006,Gerberding2015}, and the other to generate \SI{80}{\mega\Hz} sampling clocks driving the analog-digital converters (ADC) and digital clocks of the PMs. The PT calibrates the ADC sampling jitter and removes noise occurring in the \SI{80}{\mega\Hz} clock generation. Hence, its application requires high phase fidelity in the PT chain.

The primed clocks are connected to an EOM each. Note that the virtual clock $1'$, which is identical to 1, is introduced for the ease of the modeling in \cref{sec:model}. To achieve a \SI{1}{\mega\Hz} offset between the carrier-carrier beatnote and sideband-sideband beatnotes, the clocks $1'$, $2'$ and $3'$ run at \SI{2.400}{\giga\Hz}, \SI{2.401}{\giga\Hz} and \SI{2.399}{\giga\Hz}, respectively.

\begin{figure}
\begin{centering}
\includegraphics[width=8.6cm]{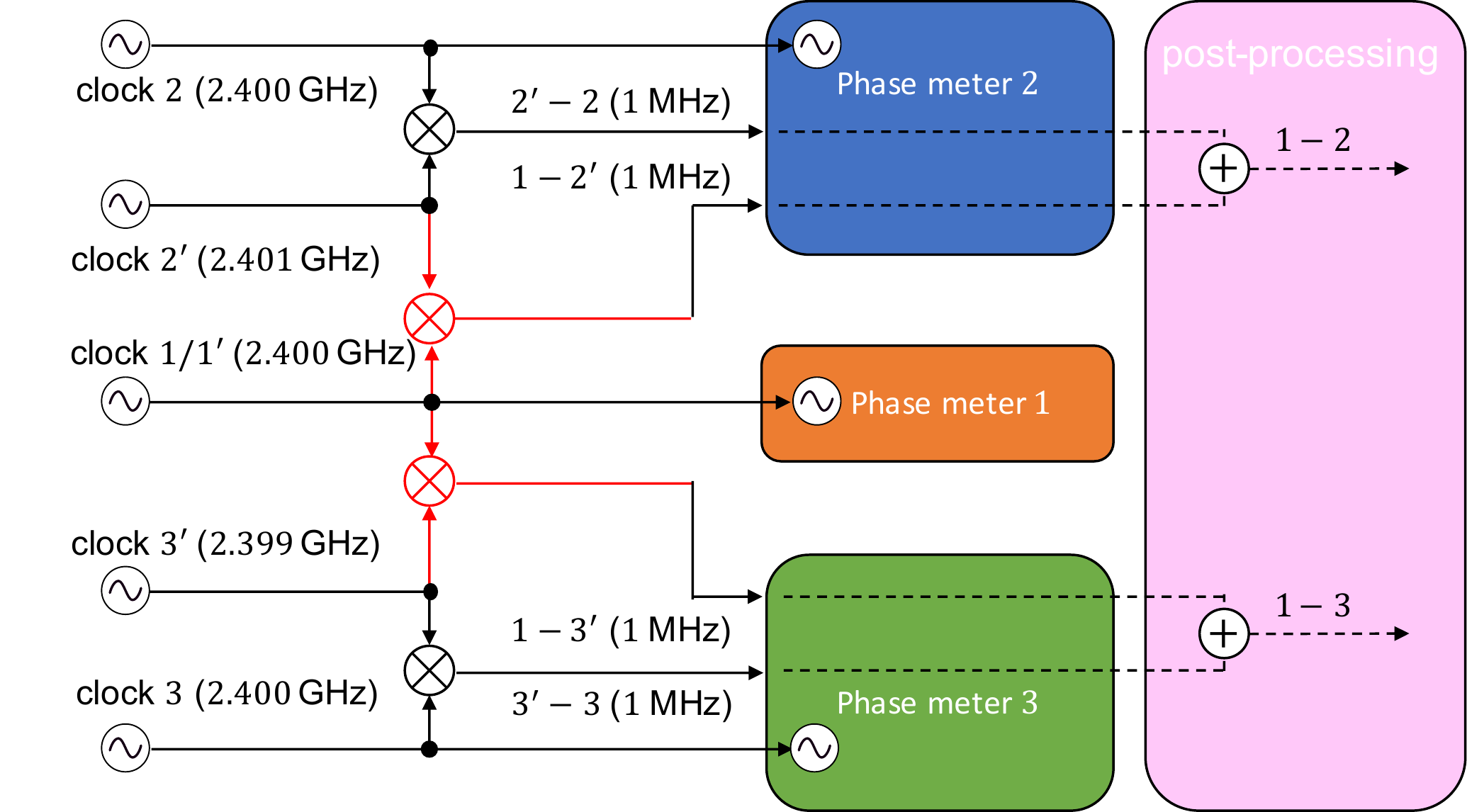}
\caption{\label{fig:clock_diagram} Conceptual diagram of the clock tone transfer. The solid-black lines, the solid-red lines and the dashed-black lines denote electrical signals, optical signals and digital data, respectively.}
\end{centering}
\end{figure}

To derive the differential clock signal between unprimed clocks driving the PMs, they need to be related to the mentioned optical sideband-sideband beatnotes. This is done by tracking electrical beatnotes between pairs of the local primed and unprimed clocks, i.e. $2'$-2 and $3'$-3. The complete clock tone transfer is conceptually summarized in \cref{fig:clock_diagram}. Through the rest of this article, clock 1 is chosen as a primary clock, while clock 2 and 3 are secondary.

In LISA, the setup will be slightly different and more symmetric. Each SC will be equipped with two EOMs, one per arm, driven by \SI{2.400}{\giga\Hz} and \SI{2.401}{\giga\Hz}. This ensures the aforementioned \SI{1}{\mega\Hz} offset. Furthermore, the different \si{\giga\Hz} signals will ultimately be derived from a single onboard \SI{10}{\mega\Hz} ultrastable oscillators (USO). The three USOs on the three SC will, however, not be actively synchronized to each other, but individually free-running.

As mentioned, the PMs operate at a sampling frequency of \SI{80}{\mega Sps} generated by the FDMs. In LISA, the data streams need to be low-pass-filtered, to avoid aliasing, and decimated in several stages to a lower data rate before being downlinked to earth. The current PM implementation in this experiment utilizes a FPGA-based cascaded integrator-comb (CIC) filter as a first stage to decimate the phase readout toward an intermediate data rate around \SI{610}{\Hz}. At these rates, more sophisticated finite impulse response (FIR) filters are applied in software to decimate further to \SI{3.4}{\Hz} as the final readout rate. Note that the LISA baseline for the final data rate was increased to \SI{4}{\Hz}. As will be described in \cref{sec:alias}, the data analysis is sensitive to insufficient low-pass filtering. The same effects will impact the data processing for LISA which in turn will require careful design of the onboard processing.

While the hexagonal optical bench and photoreceivers are hosted in a vacuum chamber with a moderate vacuum level around \SI{1}{\milli\bar}, the three PMs are mounted in housings with active temperature stabilization. The other components in \cref{fig:layout} are placed in air.

\section{Model}
\label{sec:model}
The mathematical model of the clock synchronization in this experiment can be derived by reducing the model for LISA presented in \cite{Hartwig2021}. The LISA phasemeter will be capable of providing data in either phase or frequency. While the final data format is not yet decided, this work assumes that the data is downlinked in units of frequency, which avoids the need to handle frequent phase wrapping due to the \si{\mega\Hz} beatnotes. Consequently, our experimental data is also modeled and processed using frequency units similarly to \cite{Bayle2021}, which is only integrated to compute the phase spectral density in the very end. Frequency measurements in this section are assumed to be already corrected by a PT signal to suppress ADC jitter, as described in \cite{Shaddock2006,Gerberding2015}. Moreover, it is assumed that the measured frequencies described in this section are the PM outputs, i.e. the low-rate data after the down-sampling from the \SI{80}{\mega\Hz} sampling frequency. Note that imperfections in the down-sampling process can impact the final results of our analysis, which will be studied in detail in \cref{sec:alias}. However, in this section, the in-band properties of the interferometric signals are described and thus any impact of the filters and decimation stages is neglected.

While the notation conventions used in \cite{Hartwig2021} is loosely followed, a notable difference is that this lab experiment has no need to refer to external time frames like the SC proper times. Instead, the clock $1/1'$ is chosen as a reference or primary clock and all variables are written in that clocks time frame, which will be always labeled with the index $m$ ($=m'$).

\subsection{Measurements}
The measurements available from the experiment are first described, focusing on how they are impacted by the different clocks.

The timing noise of the secondary clock $i$ relative to the primary clock $m$ is denoted as $\qt{i}(\tau)$, while the overall clock time $\tau^{\tau_m}_{i}(\tau)$ at a given reference clock time $\tau$ is modeled as 
\begin{align}
\tau^{\tau_m}_{i}(\tau) &= \tau + \delta\tau_{i}(\tau) \nonumber
\\
&= \tau + \qt{i}(\tau) + \delta\tau_{i,0}, \label{eq:tau_i_m}
\end{align}
with $\delta\tau_{i,0}$ as the constant initial time offset between the clocks. We call $\delta\tau_{i}(\tau)$ timer deviation. The superscript $\tau_m$ is to explicitly show that this functions is according to the clock $m$. The primary clock naturally satisfies $\tau^{\tau_m}_{m}(\tau) = \tau$. 

To transform variables sampled by the clock $i$ to the primary clock time frame, the inverse of \cref{eq:tau_i_m} is needed, i.e., the primary clock time given a time of the clock $i$. This can be expressed by the implicit equation
\begin{align}
\tau^{\tau_i}_{m}(\tau) &= \tau - \delta\tau_{i}(\tau^{\tau_i}_{m}(\tau)). \label{eq:tau_m_i}
\end{align}
Phase measurements $\phi^{\tau_i}$ sampled according to one of the clocks $i$ are simply shifted in time according to \cref{eq:tau_m_i},
\begin{align}
\phi^{\tau_i}(\tau) &= \phi^{\tau_m}(\tau^{\tau_i}_{m}(\tau)) \nonumber
\\
&= \phi^{\tau_m}(\tau - \delta\tau_{i}(\tau^{\tau_i}_{m}(\tau))).
\label{eq:s_t}
\end{align}

As stated above, the PM output is used in terms of frequencies, which are related to the phase by a time derivative. This gives
\begin{align}
\nu^{\tau_{i}}(\tau) &= \frac{d\phi^{\tau_i}(\tau)}{d\tau} \nonumber
\\
&=\nu^{\tau_{m}}(\tau^{\tau_i}_{m}(\tau))\cdot\frac{d\tau^{\tau_i}_{m}(\tau)}{d\tau} \nonumber
\\
&= \frac{\nu^{\tau_{m}}(\tau^{\tau_i}_{m}(\tau))}{1 + \qdt{i}(\tau^{\tau_i}_{m}(\tau))} \label{eq:nu_t}
\end{align}
for expressing a frequency measured according to the clock $i$ relative to the same frequency measured by the reference clock.

The signals in this experiment can be categorized into three types: optical carrier-carrier beatnotes, optical sideband-sideband beatnotes and the differential clock signal between the primed and unprimed clocks generated by an electrical frequency mixer. These three signals can be expressed in the primary clock time frame,
\begin{align}
\nu^{\tau_m}_{c, i'j'}(\tau) &= \nu^{\tau_m}_{j'}(\tau) - \nu^{\tau_m}_{i'}(\tau), \label{eq:nu_alpha_beta_c_t}
\\
\nu^{\tau_m}_{\rmsb, i'j'}(\tau) &= \nu^{\tau_m}_{c, i'j'}(\tau) + f_{j'}(1+\qdt{j'}(\tau)) - f_{i'}(1+\qdt{i'}(\tau)), \label{eq:nu_alpha_beta_sb_t}
\\
\nu^{\tau_m}_{\mix, i}(\tau) &= f_{i'}(1+\qdt{i'}(\tau)) - f_{i}(1+\qdt{i}(\tau)), \label{eq:nu_alpha_beta_mix_t}
\end{align}
where $f_{i}$ is the nominal frequency of the clock $i$ in the GHz regime. Note that all unprimed clocks have the same nominal frequency $f_{i} = \SI{2.400}{\giga\Hz}$. Our goal is to construct a noise-free signal combination, which in the primary clock frame would be trivially given as
\begin{align}
\Delta^{\tau_{m}}_{\rm 1PM}(\tau) = \nu^{{\tau_{m}}}_{c, 2'3'}(\tau) + \nu^{{\tau_{m}}}_{c, 1'2'}(\tau) + \nu^{{\tau_{m}}}_{c, 3'1'}(\tau) \equiv 0. \label{eq:sig3_1pm}
\end{align}
However, since the signals presented in \cref{eq:nu_alpha_beta_c_t} are recorded according to the independent clocks $i$, and thus have their expressions modified according to \cref{eq:nu_t}, all measurements need to be synchronized before $\Delta^{\tau_{m}}_{\rm 1PM}(\tau)$ can be computed.

\subsection{Clock synchronization}
Any signals sampled by the secondary clocks need to be interpolated to adjust their time stamps and rescaled to compensate the multiplicative factor in \cref{eq:nu_t}. For the former, it is first shown that a time shift by $\delta\tau_{i}(\tau)$ perfectly compensates the time stamping errors in \cref{eq:s_t},
\begin{align}
\phi^{\tau_i}(\tau + \delta\tau_i(\tau)) &= \phi^{\tau_m}(\tau^{\tau_i}_{m}(\tau + \delta\tau_i(\tau)) = \phi^{\tau_m}(\tau), \label{eq:g_interpolation}
\end{align}
since
\begin{align}
\tau^{\tau_i}_{m}(\tau + \delta\tau_{i}(\tau)) &= \tau + \delta\tau_{i}(\tau) - \delta\tau_{i}(\tau^{\tau_i}_{m}(\overbrace{\tau + \delta\tau_{i}(\tau)}^{\tau^{\tau_m}_{i}(\tau)})) = \tau. \label{eq:tau_m_i_interp}
\end{align}
Here, the following relation was used: $\tau^{\tau_i}_{m}(\tau^{\tau_m}_{i}(\tau))=\tau$.

The first step to get $\delta\tau_i(\tau)$ from our measurements is to combine \crefrange{eq:nu_alpha_beta_c_t}{eq:nu_alpha_beta_mix_t} to get a differential measurement between the primary clock $m$ and the secondary clock $i$,
\begin{align}
\rdt{\tau_{m}}{i}{m'}(\tau) = \rdt{\tau_{m}}{i}{m}(\tau) &= \frac{1}{f_{i}}\left[\left(\nu^{\tau_{m}}_{\rmsb, i'm}(\tau) - \nu^{\tau_{m}}_{c, i'm}(\tau)\right) + \nu^{\tau_{m}}_{\mix, i}(\tau)\right] \nonumber
\\
&= \qdt{m}(\tau) - \qdt{i}(\tau)\nonumber
\\
&= - \qdt{i}(\tau). \label{eq:rdt_ij}
\end{align}

The actual measurements in this combination are recorded according to the clock $i$, as shown in \cref{fig:clock_diagram}. They can be expressed by applying \cref{eq:nu_t} to $\rdt{\tau_{m}}{i}{m}(\tau)$,
\begin{align}
\rdt{\tau_{i}}{i}{m}(\tau) &= \frac{\rdt{\tau_{m}}{i}{m}(\tau^{\tau_i}_{m}(\tau))}{1 + \qdt{i}(\tau^{\tau_i}_{m}(\tau))} \nonumber
\\
&= \frac{ - \qdt{i}(\tau^{\tau_i}_{m}(\tau))}{1 + \qdt{i}(\tau^{\tau_i}_{m}(\tau))}. 
\label{eq:rde_im}
\end{align}

$\rdt{\tau_{i}}{i}{m}(\tau_{i})$ can be integrated over the measurement time, which gives, using \cref{eq:s_t,eq:nu_t},
\begin{align}
\rt{\tau_{i}}{i}{m}(\tau) &= \int^{\tau}_{0}\rdt{\tau_{i}}{i}{m}(\tau')d\tau' \nonumber
\\
&= - \qt{i}(\tau^{\tau_i}_{m}(\tau)). \label{eq:rt_im}
\end{align}
Here, our measured $\rt{\tau_{i}}{i}{m}(\tau)$ is missing the initial $\delta\tau_{i,0}$. Hence, a free parameter $\delta\hat\tau_{i,0}$ is added to \cref{eq:rt_im},
\begin{align}
\rt{\tau_{i}}{i}{m,0}(\tau) &= \rt{\tau_{i}}{i}{m}(\tau) - \delta\hat\tau_{i,0} \nonumber
\\
&= - \qt{i}(\tau^{\tau_i}_{m}(\tau)) - \delta\hat\tau_{i,0} \nonumber
\\
&\approx - \delta\tau_{i}(\tau^{\tau_i}_{m}(\tau)),
\label{eq:rt0_im}
\end{align}
where the last approximation is only realized after fitting the correct value for $\delta\hat\tau_{i,0}$.

This gives us $\delta\tau_{i}$, but still evaluated at $\tau^{\tau_i}_{m}(\tau)$. The timestamp can be adjusted by numerically solving the following nested equation,
\begin{align}
\delta\tau_{i}(\tau) &= \delta\tau_{i}(\tau^{\tau_i}_{m}(\tau + \delta\tau_{i}(\tau))) \nonumber
\\
&= - \rt{\tau_{i}}{i}{m,0}(\tau + \delta\tau_{i}(\tau)). 
\label{eq:timer_deviation_from_rt0_im}
\end{align}
Note that the final results in this experiment do not significantly change when using $\rt{\tau_{i}}{i}{m,0}(\tau)$ instead of $\delta\tau_{i}(\tau)$ for the  interpolation. The reason is that the sub-ppm frequency offsets between the clocks result in a negligible timing error over the typical measurement times in the lab.
Nevertheless, the more exact expression given in \cref{eq:timer_deviation_from_rt0_im} will be used to time-shift the measurements to the primary clock frame. In addition, it is necessary to undo the frequency scaling in \cref{eq:nu_t} by applying a multiplicative factor. In total, the following is computed,
\begin{align}
\tilde{\nu}^{\tau_{i}}_{c,i'j'}(\tau) &= \frac{\nu^{{\tau_{i}}}_{c,i'j'}(\tau+\delta\tau_{i}(\tau))}{1+\rdt{\tau_{i}}{i}{m}(\tau+\delta\tau_{i}(\tau))} \approx \nu^{{\tau_{m}}}_{c,i'j'}(\tau). 
\label{eq:tilde_nu_c}
\end{align}

The final signal combination can be now formed,
\begin{align}
\Delta_{\rm 3PM}(\tau; \delta\hat\tau_{i,0}) &= \nu^{\tau_{1}}_{c, 2'3'}(\tau) + \tilde{\nu}^{\tau_{2}}_{c, 1'2'}(\tau) + \tilde{\nu}^{\tau_{3}}_{c, 3'1'}(\tau) \approx  \Delta^{\tau_{m}}_{\rm 1PM}(\tau), \label{eq:sig3_3pm}
\end{align}
which reduces to the noise-free \cref{eq:sig3_1pm} after $\delta\hat\tau_{i,0}$ is fitted to the correct value of $\delta\tau_{i,0}$.

\section{Aliasing Effect}
\label{sec:alias}
Laser frequency noises at high Fourier frequencies are folded into the observation band due to aliasing in the decimation stages. Hence, the clock synchronization needs to take into account this frequency regime by utilizing carefully designed antialiasing filters.  As time-stamping and sampling operations do not commute (see below), the residuals of aliased frequency noise will otherwise spoil the measurements. A brief summary of this will be given in the following, while a detailed model is currently under development by Staab {\it et al}.

A decimation stage can be considered as a combination of an antialiasing filter and down-sampling. This process is expressed introducing a sampling operator $\S$ and a filter operator $\F$,
\begin{align}
\nu_{S}(\tau) &= \S\F\nu(\tau'),
\label{eq:S_define}
\end{align}
where $\nu_{S}(\tau)$ is a measured frequency after filtering and down-sampling. The one-sided power spectrum of $\nu_{S}$ is made up of the in-band contribution and any folded/aliased power which results from down-sampling,
\begin{align}
S_{\nu_{S}}(f) &= \sum_{k=0}^\infty \tilde{\F}(f)S^{(k)}_{\nu}(f).
\label{eq:S_power_spectrum}
\end{align}
where $\tilde{\F}(f)$ is the modulus squared of the filter's transfer function. $S_\nu^{(k)}(f)$ denotes the $k^{\mathrm{th}}$ alias which is given by

\begin{align}
S_{\nu}^{(k)}(f) &= 
\begin{cases*}
S_{\nu}(n[k]f_s + f), \hspace{2mm} n[k] = \frac{k}{2} & if $k$ is even\\
S_{\nu}(n[k]f_s - f), \hspace{2mm} n[k] = \frac{k+1}{2} & if $k$ is odd
\end{cases*}
\label{eq:S_k_power_spectrum}
\end{align}

Using the introduced formalism, a split measurement, i.e. the difference between the same signals measured by independent two PMs, is written with the adjustment of one of the time stamps in postprocessing, 
\begin{align}
y(\tau) &= \S\F\nu(\tau') - \Tinv{i}\S\F\T{i}\nu(\tau') \nonumber
\\
&= \Tinv{i}\left( [\T{i}, \S]\F + \S[\T{i}, \F] \right) \nu(\tau'), \label{eq:S_T_incommutativity}
\end{align}
where the time-stamping operator $\T{i}$ and its inverse operator $\Tinv{i}$ are introduced. They represent the time shifts due to the timer deviation between the primary and secondary clocks and the compensation for it by time-shifting, respectively. Notice that a timer deviation was assumed to be constant here, hence, $\qdt{i}(\tau')$, which generally appears in a denominator like \cref{eq:nu_t}, was neglected. The first term shows the commutator between time-stamping and sampling, which is the focus of the rest of this section. The second term is the one between time-stamping and filtering, which is called flexing-filtering coupling \cite{Bayle2019} and described in \cref{sec:flexing_filtering}.

The $k$-th contribution to the total amplitude spectrum density (ASD) is derived by taking a square root of the ensemble average of the squared modulus of the Fourier transform $\mathcal{F}$, 
\begin{align}
\tilde{Y}^{(k)}(f) &= \sqrt{\left\langle\left|\mathcal{F}\left[ \Tinv{i}[\T{i}, \S]\F \nu(\tau') \right] \right|^2 \right\rangle } \nonumber
\\
&=
\sqrt{\tilde{\F}(f)S_{\nu}^{(k)}(f)}\cdot\left| \e^{-i2\pi f\delta\tau_{i}} - \e^{-i2\pi (n[k]f_{s} + f)\delta\tau_{i}} \right|
\nonumber
\\
&= 
\sqrt{\tilde{\F}(f)S_{\nu}^{(k)}(f)} \cdot 2\left|\sin(\pi n[k] f_s \delta\tau_{i}) \right|.
\label{eq:split_amp_spectrum}
\end{align}
This shows $\T{i}$ and $\S$ does commute for power below the new Nyquist frequency but does not for all aliased power, i.e. when $k\neq0$.

Up to here, a constant timer deviation $\delta\tau_{i}$ was assumed to derive the ASD of the commutator. Our actual clocks have an almost constant frequency offset over a lab measurement time, such that the timer deviations are time-dependent and almost linear in time. Hence, if the timer deviation varies more than $O(1/f_s)$, the sinusoidal factor is averaged because of the phase scanning. In this case, after such an averaging, \cref{eq:S_T_incommutativity} reduces to,
\begin{align}
\tilde{Y}^{(k)}(f) &\approx \sqrt{\tilde{\F}(f)S_{\nu}^{(k)}(f)} \cdot \sqrt{2}. \label{eq:split_amp_spectrum_ave}
\end{align}

This model of the aliasing effect can be applied to the three-signal measurement,
\begin{align}
\Delta'_{\rm 3PM}(\tau) &= \S\F\nu_{c, 2'3'}(\tau') + \Tinv{2}\S\F\T{2}\nu_{c, 1'2'}(\tau') + \Tinv{3}\S\F\T{3}\nu_{c, 3'1'}(\tau') \nonumber
\\
&= - \Tinv{2} [\T{2}, \S] \F \nu_{c, 1'2'}(\tau') - \Tinv{3} [\T{3}, \S] \F \nu_{c, 3'1'}(\tau'), 
\label{eq:sig3_alias}
\end{align}
where all detailed descriptions of clock synchronizations provided in \cref{sec:model} are omitted and the perfect three-signal test without any additional noise was assumed in the second line, i.e. $\nu_{c, 2'3'}(\tau') = - \nu_{c, 1'2'}(\tau') - \nu_{c, 3'1'}(\tau')$. 

\cref{sec:results} specifically demonstrates the aliasing effect due to the CIC decimation stage mentioned in \cref{sec:setup}. The filter operator described in this section corresponds to its integration stage prior to down-sampling. This is followed by the comb stage, whose transfer function can be simply applied from the left side of \cref{eq:split_amp_spectrum_ave}.

\section{Results}
\label{sec:results}

\begin{figure*}
\begin{centering}
\includegraphics[width=17.2cm]{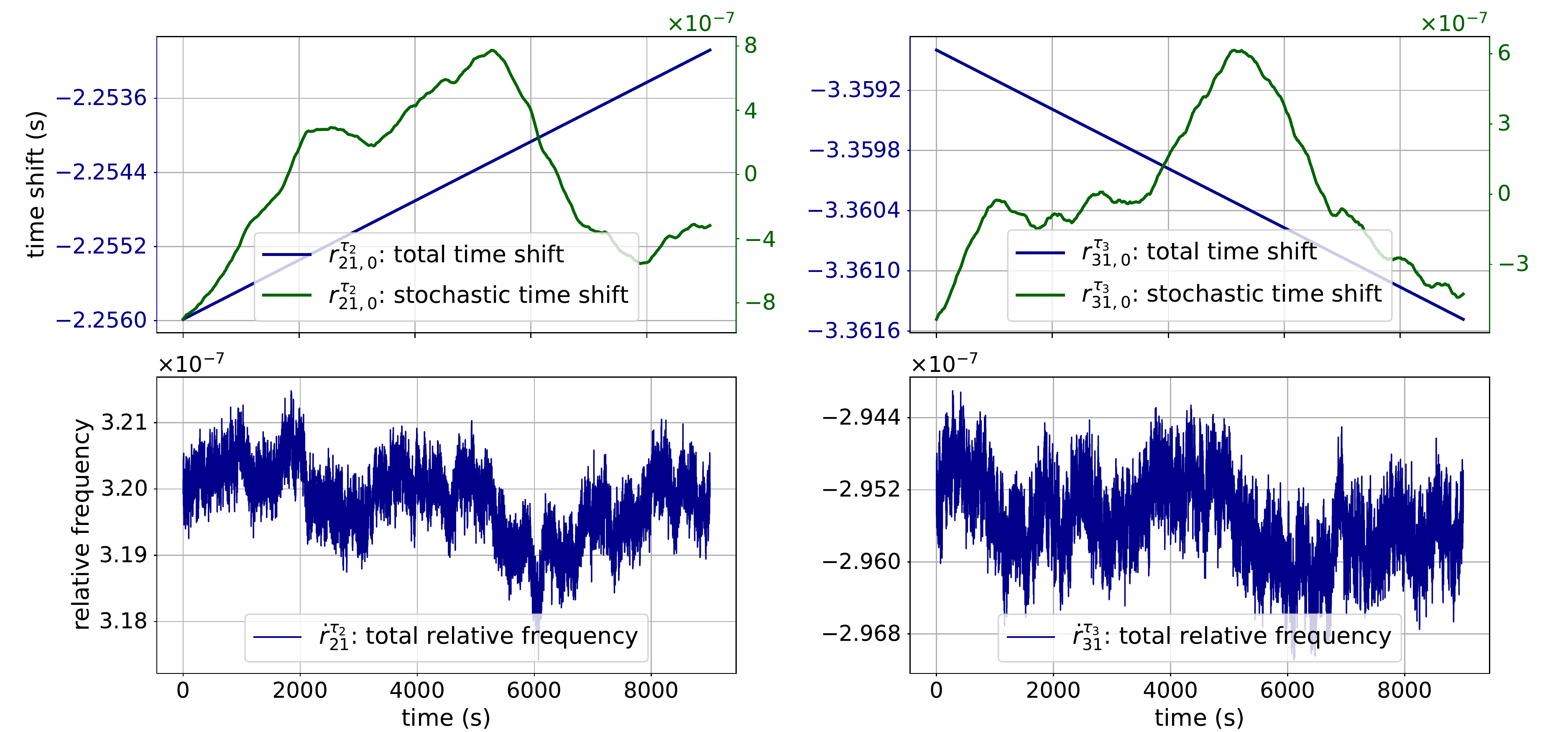}
\caption{\label{fig:diff_clock_t} Two differential clock signals. Left:  clocks 1-2. Right: clocks 1-3. The total relative frequency offsets were measured (bottom) and the mean values were estimated at \SI{0.320}{ppm} and \SI{0.296}{ppm}. These were integrated to compute the clock time shifts (top). The top panel shows the detrended time shifts (green), which shows their stochastic components, together with the total time shifts (blue).}
\end{centering}
\end{figure*}

Using the setup in \cref{sec:setup}, the clock synchronization based on the model provided in \cref{sec:model,sec:alias} was experimentally demonstrated.

As shown in \cref{fig:layout}, the two secondary lasers were locked to the primary laser with fixed \si{\mega\Hz} offsets. On top, LISA-like frequency noises was added at the lock error point. The \si{\mega\Hz} beatnote frequencies of $\nu_{c, 1'2'}$, $\nu_{c, 2'3'}$ and $\nu_{c, 3'1'}$ were chosen as follows: 23.3, 6.6 and 16.7\,\si{\mega\Hz}. The white frequency noise of the beatnotes was set to \SI{60}{\Hz\per\sqrt{\rm \Hz}}, mimicking current noise allocations for LISA. All complementary photoreciever signals were fed to PM 1. The associated three-signal combination is, hence, measured with a single clock and serves as a measurement of the testbed sensitivity for the clock synchronization, i.e. $\Delta^{\tau_{m}}_{\rm 1PM}(\tau)$ in \cref{eq:sig3_1pm}.

The filter order of the Lagrange interpolation, to realize the time shifting described in \cref{eq:tilde_nu_c}, was 121. At both ends, 150 samples  of interpolated data were truncated. Regarding the TDIR-like optimization of the initial offsets $\delta\hat\tau_{i,0}$, the noise contribution in the final signal combination $\Delta_{\rm 3PM}(\tau; \delta\hat\tau_{i,0})$ above 0.8\,\si{Hz} was filtered out before the computation of its noise power. This was done to avoid the disturbance by the dominant interpolation error close to the Nyquist frequency.

\cref{fig:diff_clock_t} shows the two differential clock measurements from \cref{fig:clock_diagram} in both relative frequency offsets and time shifts based on the signals described by \crefrange{eq:rdt_ij}{eq:rt0_im}. The mean values of the relative frequency offsets were \SI{0.320}{ppm} and \SI{0.296}{ppm} between clocks 1-2, and 1-3, respectively. Additionally, the initial offsets $\delta\tau_{i,0}$ were derived based on TDIR-like processing: $\delta\hat\tau_{2,0} \approx 2.26$\,\si{\second} and $\delta\hat\tau_{3,0} \approx 3.36$\,\si{\second}.

\begin{figure*}
\begin{centering}
\includegraphics[width=17.2cm]{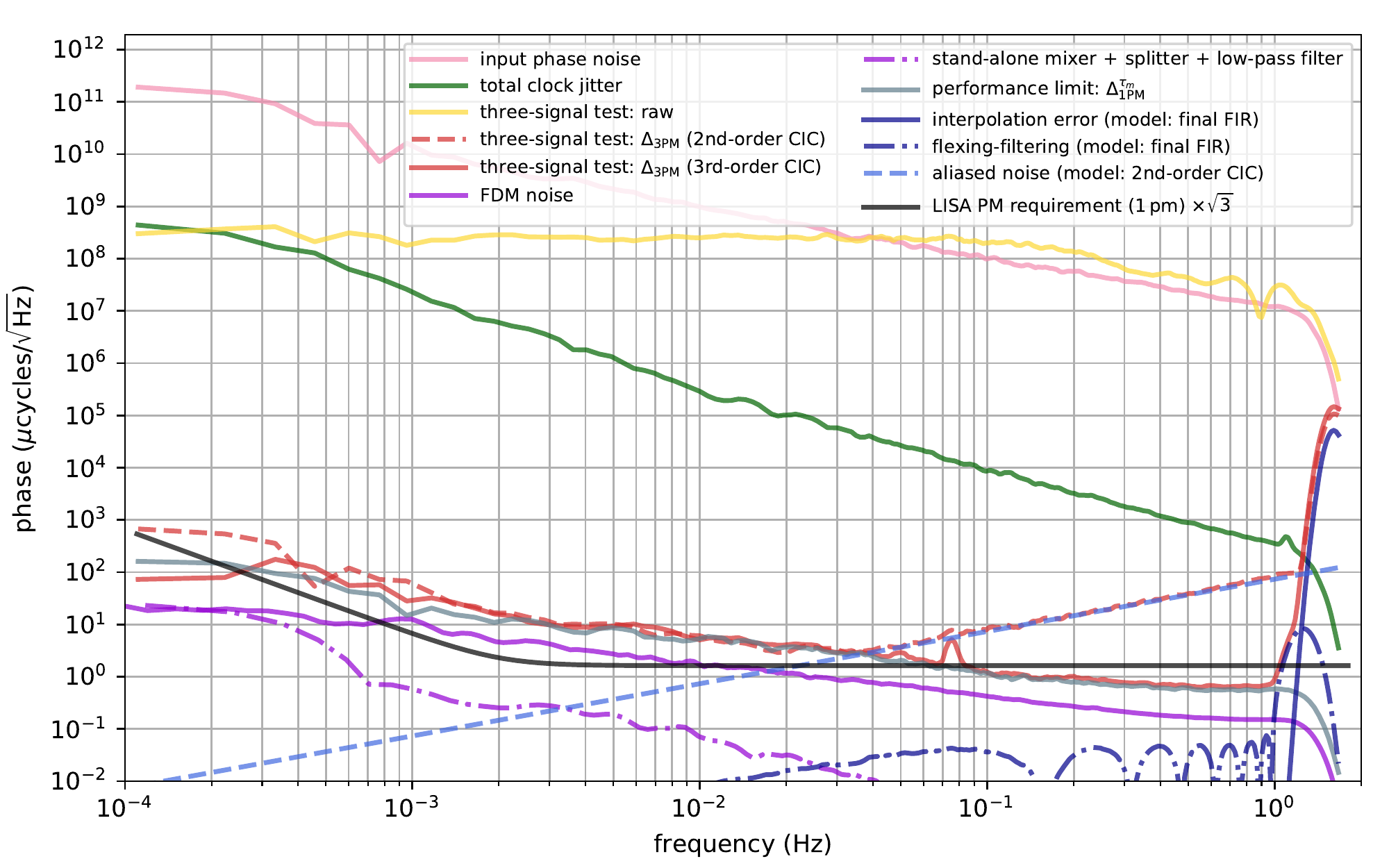}
\caption{
\label{fig:sig3_clock_sync_asd} Measurement of a three-signal test using clock synchronization. Pink shows the input phase noise due to white LISA-like frequency noise. The raw three-signal measurement with independent PMs (yellow) and the differential clock jitter (green) can be suppressed using clock synchronization (red) down to the testbed sensitivity (gray), partially reaching the LISA requirement (black). Limitations due to an insufficient 2nd order CIC filter match the model of the aliased noise (red-dashed, blue-dashed, respectively). Further limitations are due to interpolation errors (navy) and the flexing-filtering coupling (navy-dot-dashed). Noise projections of the FDM and additionally used components like stand-alone mixers are shown in violet and violet-dot-dashed, respectively. }
\end{centering}
\end{figure*}

\cref{fig:sig3_clock_sync_asd} shows the performance of the clock synchronization in phase spectrum density. Pink is one of the beatnote phase noises, which shows $1/f$ behavior due to the injected white frequency noise. Yellow shows the raw three-signal test with non-synchronized PMs. It is dominated  by effects due to the clock initial offsets. After the secondary clocks were synchronized with respect to the primary clock, see \cref{eq:sig3_3pm}, the three-signal performance (red) was suppressed with respect to the separately measured differential clock jitters (green) by 3 orders of magnitude at \SI{1}{\Hz} up to 6 orders of magnitude at \SI{0.1}{\milli\Hz}. At this state, the measurement is limited by the testbed noise monitored with a single PM (gray). It deteriorated slightly with respect to earlier measurements \cite{Schwarze2019} due to changes in the setup and is focus of ongoing further noise hunting to bridge the remaining gap to the LISA PM requirement (black) in the observation band. Comparing the noise levels between the pink and red curve allows us to characterize the achieved noise suppression in terms of an equivalent residual clock desynchronization, which in TDIR is ultimately limited by the noise floor of the testbed. This residual is estimated at a level of approximately \SI{6.15}{\nano\second}, which corresponds to \SI{1.84}{\meter} in the units of light travel time, or range. This accuracy is expected to further improve alongside the overall testbed sensitivity. For comparison, the expected PRN ranging precision is around \SI{40}{\centi\meter}\,rms \cite{Esteban2011}, with the caveat that these PRN measurements might have an undetermined bias of several meters due to cables and processing delays, which could be calibrated with TDIR.

For the regime above \SI{1}{\Hz}, two noise couplings stemming from the clock synchronization were verified. For both models, the mentioned final FIR decimation stage is significant. First, the bump close to the Nyquist frequency can be explained well by the analytical model of Lagrange interpolation errors (navy) described in \cref{sec:interpolation}. The interpolation error sharply drops down to numerical noise levels toward lower frequencies. Second, the flexing-filtering coupling (navy-dot-dashed) described in \cref{sec:flexing_filtering} limited the performance around \SI{1}{\Hz}. To gain more margin at \SI{1}{\Hz}, the FIR filter could be tuned more carefully.

As mentioned in \cref{sec:alias}, also the filter design of the CIC decimation stage has a significant impact due to aliasing effects. This effect was estimated based on the averaged model presented in  \cref{{eq:split_amp_spectrum_ave}} because the averaging is necessary as \cref{fig:diff_clock_t} shows the measured time shifts (top blue) varied by around \SI{2}{\milli\second} over a measurement time of \SI{9000}{\second}, which is longer than $1/f_s$ with $f_s$ of \SI{610}{\Hz}. Using a 2nd-order CIC, the injected frequency noise \SI{60}{\Hz\per\sqrt{\rm \Hz}} around the slow sampling rate of the CIC was aliased to the observation band. It dominates the associated three-signal measurements (red-dashed) at the upper \si{\Hz} regime according to its model (blue-dashed). In contrast, a 3rd-order CIC showed sufficient suppression and was used instead. 

Lastly, anticipating future limitations and analyzing the remaining noise margin with respect to the current testbed sensitivity, noise projections of electrical devices in the sensitive clock path were performed. More precisely, the FDM noise was measured separately and scaled to the heterodyne frequencies of this particular measurement (violet). The PT chain of this device consists of five by-2 dividers, an amplifier, a filter and a power splitter. Their in-air thermal stability notably affects the FDM performance. Besides, the total noise projection of other electrical components, namely stand-alone mixers etc. shown in \cref{fig:layout}, were again separately measured (violet-dot-dashed).

\section{Conclusion}
\label{sec:conclusion}
In this article, the experimental demonstration of intersatellite clock synchronization using a hexagonal optical bench was presented. The setup parameters are close to the current LISA baseline design. In addition to the experimental setup, data analysis techniques were assembled that are applicable to the real LISA data with minor adaptations. They are correcting for clock errors by properly time-shifting and rescaling the total frequency errors, a scheme which is currently also under investigation for LISA \cite{HartwigUnpublished}. This is in contrast to previously suggested clock correction schemes for LISA, which operate on the residual frequency fluctuations after large trends have been removed \cite{ClockHellings, ClockTinto1, ClockTinto2, Hartwig2021}.

With input frequency noise of \SI{60}{\Hz\per\sqrt{\rm \Hz}} and LISA-like heterodyne frequencies, the performance of the clock synchronization was successfully tuned down to the current testbed sensitivity, which is below \SI{1}{\pico\meter\per\sqrt{\rm\Hz}} above \SI{60}{\milli\Hz} and below \SI{10}{\pico\meter\per\sqrt{\rm\Hz}} above \SI{2}{\milli\Hz}. This presents a new benchmark performance of the LISA clock synchronization scheme. 

In addition to the shown performance, three important noise couplings were experimentally demonstrated for the first time in this context: the impact of the out-of-band frequency noise via aliasing, the interpolation error and the flexing-filtering coupling. In particular the first is necessary to constrain the filter designs for LISA by experimentally showing and characterizing the significance of out-of-band frequency noise. The interpolation error and the flexing-filtering coupling are important especially to push the upper bound of the LISA observation band to \SI{1}{\Hz} or higher.

Future investigations will include the improvement of the current testbed sensitivity to LISA PM requirement levels in the whole band. Furthermore, critical components like the FDM will be analyzed to improve their thermal stability and their coupling into the measurement performance. On longer timescales, the experimental setup will be upgraded to include more LISA features and components. As the main testbed for the LISA metrology chain, it will continue to serve as a pillar for technology development to finally enable gravitational wave detection in space.

\begin{acknowledgments}
The authors acknowledge financial support by the German Aerospace Center (DLR) with funds from the Federal Ministry of Economics and Technology (BMWi) according to a decision of the German Federal Parliament (Grants No. 50OQ0601, No. 50OQ1301, No. 50OQ1801), the European Space Agency (ESA) (Grants No. 22331/09/ NL/HB, No. 16238/10/NL/HB), the Deutsche Forschungs- gemeinschaft (DFG) Sonderforschungsbereich 1128 Relativistic Geodesy and Cluster of Excellence “QuantumFrontiers: Light and Matter at the Quantum Frontier: Foundations and Applications in Metrology” (EXC-2123, Project No. 390837967). O. H. acknowledges funding from Centre National d’Etudes Spatiales.

\end{acknowledgments}

\appendix
\section{LAGRANGE INTERPOLATION}
\label{sec:interpolation}
In postprocessing, PM data is interpolated using fractional delay filters to adjust data time stamps. We recall the analytical model of the interpolation error presented in \cite{HartwigPhD}. Time shifts are assumed to be constant in this section.

The interpolation consists of two steps: an integer delay $\T{i}^{0}$ to shift the time stamps to the nearest sample and a fractional delay $\T{i}^{\eps}$ implemented by a noncausal FIR filter. To model the interpolation based on this decomposition, the time shift $\delta\tau_i$ is also expanded, 
\begin{align}
\delta\tau_i &= \delta\tau_{i}^{0} + \delta\tau_{i}^{\eps},
\label{eq:time_shift_decomposition}
\end{align}
where $\delta\tau_{i}^{0}=N/f_s$ by definition. $f_s$ is the sampling frequency and $N$ is an integer.

To estimate the error in frequency domain, the Fourier transform of the following expression needs to be derived, 
\begin{align}
\delta\T{i}x(t) = \left[\T{i}^{\eps}\T{i}^{0} - \mathcal{T}_{i}\right]x(t), 
\label{eq:delta_T}
\end{align}
where $x(t)$ is a given data in time and $\mathcal{T}_{i}$ is the perfect operator.

The Fourier transform of each operator reads, 
\begin{align}
\mathcal{F}\left[\T{i}^{0}x(t)\right](\omega) &= \e^{i\omega \delta\tau_{i}^{0}}\tilde{x}(\omega), \label{eq:fourier_int_delay}
\\
\mathcal{F}\left[\T{i}^{\eps}x(t)\right](\omega) &= \sum^{p}_{k=-p+1}c^{\eps}_{k}\e^{i\omega k/f_s}\tilde{x}(\omega), \label{eq:fourier_eps_delay}
\\
\mathcal{F}\left[\mathcal{T}_{i}x(t)\right](\omega) &= \e^{i\omega \left(\delta\tau_{i}^{0}+\delta\tau_{i}^{\eps}\right)}\tilde{x}(\omega), \label{eq:fourier_perfect_delay}
\end{align}
where $p=(a+1)/2$ with $a$ as the filter order and $c^{\eps}_{k}$ is a filter coefficient.

Combining these equations, the Fourier transform of \cref{eq:delta_T} is derived,
\begin{align}
\mathcal{F}\left[\delta \mathbf{T}_{i}x(t)\right](\omega) &= \e^{i\omega \delta\tau_{i}^{0}}\left[\sum^{p}_{k=-p+1}\left(c^{\eps}_{k}\e^{i\omega k/f_s}\right) - \e^{i\omega \delta\tau_{i}^{\eps}}\right]\tilde{x}(\omega). \label{eq:fourier_T_error}
\end{align}

After all, the interpolation error in amplitude spectral density $\delta \tilde{\mathbf{T}}_{i}(\omega)$ is computed, 
\begin{align}
\delta \tilde{\mathbf{T}}_{i}(\omega) &= \left|\sum^{p}_{k=-p+1}\left(c^{\eps}_{k}\e^{i\omega k/f_s}\right) - \e^{i\omega \delta\tau_{i}^{\eps}}\right|. 
\label{eq:T_error_asd}
\end{align}

\section{FLEXING FILTERING}
\label{sec:flexing_filtering}
The model of the flexing filtering coupling presented in \cite{Bayle2019} is adapted to our case where only a single time-stamping operator exists in \cref{eq:S_T_incommutativity}. 

The Fourier transform of time-shifted data $\T{i}x(t)$ can be generally expressed,
\begin{align}
\mathcal{F}\left[\T{i}x(t)\right](\omega) &= \int^{\infty}_{-\infty}x(t-\delta\tau_i(t))dt \nonumber
\\
&= \frac{1}{1-\qdt{i}}\exp\left(-i\omega\frac{\delta\tau_{i,0}}{1-\qdt{i}}\right)\tilde{x}\left(\frac{\omega}{1-\qdt{i}}\right),
\label{eq:fourier_shifted}
\end{align}
where the timer deviation $\delta\tau_i(t)$ was assumed to be a linear function of time, i.e. $\delta\tau_i(t) = \qdt{i}t + \delta\tau_{i,0}$.

Each term of the commutator between time-stamping and filtering reads,
\begin{align}
\mathcal{F}\left[\T{i}\F x(t)\right](\omega) &= \frac{1}{1-\qdt{i}}\exp\left(-i\omega\frac{\delta\tau_{i,0}}{1-\qdt{i}}\right)\tilde{x}\left(\frac{\omega}{1-\qdt{i}}\right)\tilde{\F}(\omega), 
\label{eq:fourier_TFx}
\\
\mathcal{F}\left[\F\T{i} x(t)\right](\omega) &= \frac{1}{1-\qdt{i}}\exp\left(-i\omega\frac{\delta\tau_{i,0}}{1-\qdt{i}}\right)\tilde{x}\left(\frac{\omega}{1-\qdt{i}}\right)\tilde{\F}\left(\frac{\omega}{1-\qdt{i}}\right).
\label{eq:fourier_FTx}
\end{align}

Using \cref{eq:fourier_TFx,eq:fourier_FTx}, the flexing filtering coupling to the first order of $\qdt{i}$ is derived,
\begin{align}
\mathcal{F}\left[[\T{i}, \F] x(t)\right](\omega) &\approx \omega\qdt{i}\exp\left(-i\omega\delta\tau_{i,0}\right)\tilde{x}\left(\omega\right)\frac{d\tilde{\F}(\omega)}{d\omega}. 
\label{eq:fourier_ff}
\end{align}

Hence, a filter design couples to a phase measurement via its frequency derivative. This implies that this flexing filtering coupling is dominant close to the Nyquist frequency of a particular decimation stage.

\bibliography{hex}

\begin{thebibliography}{23}%
\makeatletter
\providecommand \@ifxundefined [1]{%
 \@ifx{#1\undefined}
}%
\providecommand \@ifnum [1]{%
 \ifnum #1\expandafter \@firstoftwo
 \else \expandafter \@secondoftwo
 \fi
}%
\providecommand \@ifx [1]{%
 \ifx #1\expandafter \@firstoftwo
 \else \expandafter \@secondoftwo
 \fi
}%
\providecommand \natexlab [1]{#1}%
\providecommand \enquote  [1]{``#1''}%
\providecommand \bibnamefont  [1]{#1}%
\providecommand \bibfnamefont [1]{#1}%
\providecommand \citenamefont [1]{#1}%
\providecommand \href@noop [0]{\@secondoftwo}%
\providecommand \href [0]{\begingroup \@sanitize@url \@href}%
\providecommand \@href[1]{\@@startlink{#1}\@@href}%
\providecommand \@@href[1]{\endgroup#1\@@endlink}%
\providecommand \@sanitize@url [0]{\catcode `\\12\catcode `\$12\catcode
  `\&12\catcode `\#12\catcode `\^12\catcode `\_12\catcode `\%12\relax}%
\providecommand \@@startlink[1]{}%
\providecommand \@@endlink[0]{}%
\providecommand \url  [0]{\begingroup\@sanitize@url \@url }%
\providecommand \@url [1]{\endgroup\@href {#1}{\urlprefix }}%
\providecommand \urlprefix  [0]{URL }%
\providecommand \Eprint [0]{\href }%
\providecommand \doibase [0]{https://doi.org/}%
\providecommand \selectlanguage [0]{\@gobble}%
\providecommand \bibinfo  [0]{\@secondoftwo}%
\providecommand \bibfield  [0]{\@secondoftwo}%
\providecommand \translation [1]{[#1]}%
\providecommand \BibitemOpen [0]{}%
\providecommand \bibitemStop [0]{}%
\providecommand \bibitemNoStop [0]{.\EOS\space}%
\providecommand \EOS [0]{\spacefactor3000\relax}%
\providecommand \BibitemShut  [1]{\csname bibitem#1\endcsname}%
\let\auto@bib@innerbib\@empty
\bibitem [{\citenamefont {Abbott}\ \emph {et~al.}(2016)\citenamefont {Abbott}
  \emph {et~al.}}]{GW150914}%
  \BibitemOpen
  \bibfield  {author} {\bibinfo {author} {\bibfnamefont {B.~P.}\ \bibnamefont
  {Abbott}} \emph {et~al.} (\bibinfo {collaboration} {LIGO and Virgo
  Collaboration}),\ }\bibfield  {title} {\bibinfo {title} {Observation of
  gravitational waves from a binary black hole merger},\ }\bibfield  {journal}
  {\bibinfo  {journal} {Physical Review Letters}\ }\textbf {\bibinfo {volume}
  {116}},\ \href {https://doi.org/10.1103/physrevlett.116.061102}
  {10.1103/physrevlett.116.061102} (\bibinfo {year} {2016})\BibitemShut
  {NoStop}%
\bibitem [{\citenamefont {Tinto}\ and\ \citenamefont
  {Armstrong}(1999)}]{Tinto1999}%
  \BibitemOpen
  \bibfield  {author} {\bibinfo {author} {\bibfnamefont {M.}~\bibnamefont
  {Tinto}}\ and\ \bibinfo {author} {\bibfnamefont {J.~W.}\ \bibnamefont
  {Armstrong}},\ }\bibfield  {title} {\bibinfo {title} {Cancellation of laser
  noise in an unequal-arm interferometer detector of gravitational radiation},\
  }\href {https://doi.org/10.1103/PhysRevD.59.102003} {\bibfield  {journal}
  {\bibinfo  {journal} {Phys. Rev. D}\ }\textbf {\bibinfo {volume} {59}},\
  \bibinfo {pages} {102003} (\bibinfo {year} {1999})}\BibitemShut {NoStop}%
\bibitem [{\citenamefont {Weaver}\ \emph {et~al.}(2010)\citenamefont {Weaver},
  \citenamefont {Garstecki},\ and\ \citenamefont {Reynolds}}]{Weaver2010}%
  \BibitemOpen
  \bibfield  {author} {\bibinfo {author} {\bibfnamefont {G.}~\bibnamefont
  {Weaver}}, \bibinfo {author} {\bibfnamefont {J.}~\bibnamefont {Garstecki}},\
  and\ \bibinfo {author} {\bibfnamefont {S.}~\bibnamefont {Reynolds}},\
  }\bibfield  {title} {\bibinfo {title} {The performance of ultra-stable
  oscillators for the gravity recovery and interior laboratory (grail)},\ }in\
  \href@noop {} {\emph {\bibinfo {booktitle} {Proceedings of the 42nd Annual
  Precise Time and Time Interval (PTTI) Systems and Applications Meeting
  2010}}}\ (\bibinfo {year} {2010})\ \bibinfo {note}
  {\url{https://www.ion.org/publications/abstract.cfm?articleID=10744}}\BibitemShut
  {NoStop}%
\bibitem [{\citenamefont {Tinto}\ \emph
  {et~al.}(2002{\natexlab{a}})\citenamefont {Tinto}, \citenamefont
  {Estabrook},\ and\ \citenamefont {Armstrong}}]{Tinto2001}%
  \BibitemOpen
  \bibfield  {author} {\bibinfo {author} {\bibfnamefont {M.}~\bibnamefont
  {Tinto}}, \bibinfo {author} {\bibfnamefont {F.~B.}\ \bibnamefont
  {Estabrook}},\ and\ \bibinfo {author} {\bibfnamefont {J.~W.}\ \bibnamefont
  {Armstrong}},\ }\bibfield  {title} {\bibinfo {title} {Time-delay
  interferometry for {LISA}},\ }\href
  {https://doi.org/10.1103/PhysRevD.65.082003} {\bibfield  {journal} {\bibinfo
  {journal} {Phys. Rev. D}\ }\textbf {\bibinfo {volume} {65}},\ \bibinfo
  {pages} {082003} (\bibinfo {year} {2002}{\natexlab{a}})}\BibitemShut
  {NoStop}%
\bibitem [{\citenamefont {Hartwig}\ and\ \citenamefont
  {Bayle}(2021)}]{Hartwig2021}%
  \BibitemOpen
  \bibfield  {author} {\bibinfo {author} {\bibfnamefont {O.}~\bibnamefont
  {Hartwig}}\ and\ \bibinfo {author} {\bibfnamefont {J.-B.}\ \bibnamefont
  {Bayle}},\ }\bibfield  {title} {\bibinfo {title} {Clock-jitter reduction in
  {LISA} time-delay interferometry combinations},\ }\href
  {https://doi.org/10.1103/PhysRevD.103.123027} {\bibfield  {journal} {\bibinfo
   {journal} {Phys. Rev. D}\ }\textbf {\bibinfo {volume} {103}},\ \bibinfo
  {pages} {123027} (\bibinfo {year} {2021})}\BibitemShut {NoStop}%
\bibitem [{\citenamefont {Bender}\ \emph {et~al.}(1998)\citenamefont {Bender},
  \citenamefont {Danzmann} \emph {et~al.}}]{PrePhaseA}%
  \BibitemOpen
  \bibfield  {author} {\bibinfo {author} {\bibfnamefont {P.}~\bibnamefont
  {Bender}}, \bibinfo {author} {\bibfnamefont {K.}~\bibnamefont {Danzmann}},
  \emph {et~al.} (\bibinfo {collaboration} {the {LISA} Study Team}),\
  }\href@noop {} {\bibinfo {title} {{LISA}. laser interferometer space antenna
  for the detection and observation of gravitational waves. an international
  project in the field of fundamental physics in space}} (\bibinfo {year}
  {1998}),\ \bibinfo {note}
  {\url{https://pure.mpg.de/pubman/faces/ViewItemFullPage.jsp?itemId=item_52082}}\BibitemShut
  {NoStop}%
\bibitem [{\citenamefont {Heinzel}\ \emph {et~al.}(2011)\citenamefont
  {Heinzel}, \citenamefont {Esteban}, \citenamefont {Barke}, \citenamefont
  {Otto}, \citenamefont {Wang}, \citenamefont {Garcia},\ and\ \citenamefont
  {Danzmann}}]{Heinzel2011}%
  \BibitemOpen
  \bibfield  {author} {\bibinfo {author} {\bibfnamefont {G.}~\bibnamefont
  {Heinzel}}, \bibinfo {author} {\bibfnamefont {J.~J.}\ \bibnamefont
  {Esteban}}, \bibinfo {author} {\bibfnamefont {S.}~\bibnamefont {Barke}},
  \bibinfo {author} {\bibfnamefont {M.}~\bibnamefont {Otto}}, \bibinfo {author}
  {\bibfnamefont {Y.}~\bibnamefont {Wang}}, \bibinfo {author} {\bibfnamefont
  {A.~F.}\ \bibnamefont {Garcia}},\ and\ \bibinfo {author} {\bibfnamefont
  {K.}~\bibnamefont {Danzmann}},\ }\bibfield  {title} {\bibinfo {title}
  {Auxiliary functions of the {LISA} laser link: ranging, clock noise transfer
  and data communication},\ }\href
  {https://doi.org/10.1088/0264-9381/28/9/094008} {\bibfield  {journal}
  {\bibinfo  {journal} {Classical and Quantum Gravity}\ }\textbf {\bibinfo
  {volume} {28}},\ \bibinfo {pages} {094008} (\bibinfo {year}
  {2011})}\BibitemShut {NoStop}%
\bibitem [{\citenamefont {Esteban}\ \emph {et~al.}(2011)\citenamefont
  {Esteban}, \citenamefont {Garc\'{i}a}, \citenamefont {Barke}, \citenamefont
  {Peinado}, \citenamefont {Cervantes}, \citenamefont {Bykov}, \citenamefont
  {Heinzel},\ and\ \citenamefont {Danzmann}}]{Esteban2011}%
  \BibitemOpen
  \bibfield  {author} {\bibinfo {author} {\bibfnamefont {J.~J.}\ \bibnamefont
  {Esteban}}, \bibinfo {author} {\bibfnamefont {A.~F.}\ \bibnamefont
  {Garc\'{i}a}}, \bibinfo {author} {\bibfnamefont {S.}~\bibnamefont {Barke}},
  \bibinfo {author} {\bibfnamefont {A.~M.}\ \bibnamefont {Peinado}}, \bibinfo
  {author} {\bibfnamefont {F.~G.}\ \bibnamefont {Cervantes}}, \bibinfo {author}
  {\bibfnamefont {I.}~\bibnamefont {Bykov}}, \bibinfo {author} {\bibfnamefont
  {G.}~\bibnamefont {Heinzel}},\ and\ \bibinfo {author} {\bibfnamefont
  {K.}~\bibnamefont {Danzmann}},\ }\bibfield  {title} {\bibinfo {title}
  {Experimental demonstration of weak-light laser ranging and data
  communication for lisa},\ }\href {https://doi.org/10.1364/OE.19.015937}
  {\bibfield  {journal} {\bibinfo  {journal} {Opt. Express}\ }\textbf {\bibinfo
  {volume} {19}},\ \bibinfo {pages} {15937} (\bibinfo {year}
  {2011})}\BibitemShut {NoStop}%
\bibitem [{\citenamefont {Tinto}\ \emph {et~al.}(2005)\citenamefont {Tinto},
  \citenamefont {Vallisneri},\ and\ \citenamefont {Armstrong}}]{Tinto2005}%
  \BibitemOpen
  \bibfield  {author} {\bibinfo {author} {\bibfnamefont {M.}~\bibnamefont
  {Tinto}}, \bibinfo {author} {\bibfnamefont {M.}~\bibnamefont {Vallisneri}},\
  and\ \bibinfo {author} {\bibfnamefont {J.~W.}\ \bibnamefont {Armstrong}},\
  }\bibfield  {title} {\bibinfo {title} {Time-delay interferometric ranging for
  space-borne gravitational-wave detectors},\ }\href
  {https://doi.org/10.1103/PhysRevD.71.041101} {\bibfield  {journal} {\bibinfo
  {journal} {Phys. Rev. D}\ }\textbf {\bibinfo {volume} {71}},\ \bibinfo
  {pages} {041101(R)} (\bibinfo {year} {2005})}\BibitemShut {NoStop}%
\bibitem [{\citenamefont {Danzmann}\ \emph {et~al.}(2017)\citenamefont
  {Danzmann}, \citenamefont {Amaro-Seoane}, \citenamefont {Audley},
  \citenamefont {Babak}, \citenamefont {Baker}, \citenamefont {Barausse},
  \citenamefont {Bender}, \citenamefont {Berti}, \citenamefont {Binetruy},
  \citenamefont {Born} \emph {et~al.}}]{LISA_L3}%
  \BibitemOpen
  \bibfield  {author} {\bibinfo {author} {\bibfnamefont {K.}~\bibnamefont
  {Danzmann}}, \bibinfo {author} {\bibfnamefont {P.}~\bibnamefont
  {Amaro-Seoane}}, \bibinfo {author} {\bibfnamefont {H.}~\bibnamefont
  {Audley}}, \bibinfo {author} {\bibfnamefont {S.}~\bibnamefont {Babak}},
  \bibinfo {author} {\bibfnamefont {J.}~\bibnamefont {Baker}}, \bibinfo
  {author} {\bibfnamefont {E.}~\bibnamefont {Barausse}}, \bibinfo {author}
  {\bibfnamefont {P.}~\bibnamefont {Bender}}, \bibinfo {author} {\bibfnamefont
  {E.}~\bibnamefont {Berti}}, \bibinfo {author} {\bibfnamefont
  {P.}~\bibnamefont {Binetruy}}, \bibinfo {author} {\bibfnamefont
  {M.}~\bibnamefont {Born}}, \emph {et~al.},\ }\href@noop {} {\bibinfo {title}
  {{LISA} {L3} proposal}} (\bibinfo {year} {2017}),\ \bibinfo {note}
  {\url{https://www.elisascience.org/articles/lisa-mission/lisa-mission-proposal-l3}}\BibitemShut
  {NoStop}%
\bibitem [{\citenamefont {Schwarze}\ \emph {et~al.}(2019)\citenamefont
  {Schwarze}, \citenamefont {Fern\'andez~Barranco}, \citenamefont {Penkert},
  \citenamefont {Kaufer}, \citenamefont {Gerberding},\ and\ \citenamefont
  {Heinzel}}]{Schwarze2019}%
  \BibitemOpen
  \bibfield  {author} {\bibinfo {author} {\bibfnamefont {T.~S.}\ \bibnamefont
  {Schwarze}}, \bibinfo {author} {\bibfnamefont {G.}~\bibnamefont
  {Fern\'andez~Barranco}}, \bibinfo {author} {\bibfnamefont {D.}~\bibnamefont
  {Penkert}}, \bibinfo {author} {\bibfnamefont {M.}~\bibnamefont {Kaufer}},
  \bibinfo {author} {\bibfnamefont {O.}~\bibnamefont {Gerberding}},\ and\
  \bibinfo {author} {\bibfnamefont {G.}~\bibnamefont {Heinzel}},\ }\bibfield
  {title} {\bibinfo {title} {Picometer-stable hexagonal optical bench to verify
  {LISA} phase extraction linearity and precision},\ }\href
  {https://doi.org/10.1103/PhysRevLett.122.081104} {\bibfield  {journal}
  {\bibinfo  {journal} {Phys. Rev. Lett.}\ }\textbf {\bibinfo {volume} {122}},\
  \bibinfo {pages} {081104} (\bibinfo {year} {2019})}\BibitemShut {NoStop}%
\bibitem [{\citenamefont {de~Vine}\ \emph {et~al.}(2010)\citenamefont
  {de~Vine}, \citenamefont {Ware}, \citenamefont {McKenzie}, \citenamefont
  {Spero}, \citenamefont {Klipstein},\ and\ \citenamefont
  {Shaddock}}]{Vine2010}%
  \BibitemOpen
  \bibfield  {author} {\bibinfo {author} {\bibfnamefont {G.}~\bibnamefont
  {de~Vine}}, \bibinfo {author} {\bibfnamefont {B.}~\bibnamefont {Ware}},
  \bibinfo {author} {\bibfnamefont {K.}~\bibnamefont {McKenzie}}, \bibinfo
  {author} {\bibfnamefont {R.~E.}\ \bibnamefont {Spero}}, \bibinfo {author}
  {\bibfnamefont {W.~M.}\ \bibnamefont {Klipstein}},\ and\ \bibinfo {author}
  {\bibfnamefont {D.~A.}\ \bibnamefont {Shaddock}},\ }\bibfield  {title}
  {\bibinfo {title} {Experimental demonstration of time-delay interferometry
  for the laser interferometer space antenna},\ }\href
  {https://doi.org/10.1103/PhysRevLett.104.211103} {\bibfield  {journal}
  {\bibinfo  {journal} {Phys. Rev. Lett.}\ }\textbf {\bibinfo {volume} {104}},\
  \bibinfo {pages} {211103} (\bibinfo {year} {2010})}\BibitemShut {NoStop}%
\bibitem [{\citenamefont {Hartwig}\ \emph {et~al.}(2022)\citenamefont
  {Hartwig}, \citenamefont {Bayle}, \citenamefont {Staab}, \citenamefont
  {Hees}, \citenamefont {Lilley},\ and\ \citenamefont
  {Wolf}}]{HartwigUnpublished}%
  \BibitemOpen
  \bibfield  {author} {\bibinfo {author} {\bibfnamefont {O.}~\bibnamefont
  {Hartwig}}, \bibinfo {author} {\bibfnamefont {J.-B.}\ \bibnamefont {Bayle}},
  \bibinfo {author} {\bibfnamefont {M.}~\bibnamefont {Staab}}, \bibinfo
  {author} {\bibfnamefont {A.}~\bibnamefont {Hees}}, \bibinfo {author}
  {\bibfnamefont {M.}~\bibnamefont {Lilley}},\ and\ \bibinfo {author}
  {\bibfnamefont {P.}~\bibnamefont {Wolf}},\ }\bibfield  {title} {\bibinfo
  {title} {Time delay interferometry without clock synchronisation},\ }\Eprint
  {https://arxiv.org/abs/2202.01124} {arXiv:2202.01124 [gr-qc]}  (\bibinfo
  {year} {2022})\BibitemShut {NoStop}%
\bibitem [{\citenamefont {Schwarze}\ \emph {et~al.}(2016)\citenamefont
  {Schwarze}, \citenamefont {Barranco}, \citenamefont {Penkert}, \citenamefont
  {Gerberding}, \citenamefont {Heinzel},\ and\ \citenamefont
  {Danzmann}}]{Schwarze2016}%
  \BibitemOpen
  \bibfield  {author} {\bibinfo {author} {\bibfnamefont {T.~S.}\ \bibnamefont
  {Schwarze}}, \bibinfo {author} {\bibfnamefont {G.~F.}\ \bibnamefont
  {Barranco}}, \bibinfo {author} {\bibfnamefont {D.}~\bibnamefont {Penkert}},
  \bibinfo {author} {\bibfnamefont {O.}~\bibnamefont {Gerberding}}, \bibinfo
  {author} {\bibfnamefont {G.}~\bibnamefont {Heinzel}},\ and\ \bibinfo {author}
  {\bibfnamefont {K.}~\bibnamefont {Danzmann}},\ }\bibfield  {title} {\bibinfo
  {title} {Optical testbed for the {LISA} phasemeter},\ }in\ \href
  {https://iopscience.iop.org/article/10.1088/1742-6596/716/1/012004} {\emph
  {\bibinfo {booktitle} {J. Phys. Conf. Ser.}}}\ (\bibinfo {year}
  {2016})\BibitemShut {NoStop}%
\bibitem [{\citenamefont {Shaddock}\ \emph {et~al.}(2006)\citenamefont
  {Shaddock}, \citenamefont {Ware}, \citenamefont {Halverson}, \citenamefont
  {Spero},\ and\ \citenamefont {Klipstein}}]{Shaddock2006}%
  \BibitemOpen
  \bibfield  {author} {\bibinfo {author} {\bibfnamefont {D.}~\bibnamefont
  {Shaddock}}, \bibinfo {author} {\bibfnamefont {B.}~\bibnamefont {Ware}},
  \bibinfo {author} {\bibfnamefont {P.~G.}\ \bibnamefont {Halverson}}, \bibinfo
  {author} {\bibfnamefont {R.~E.}\ \bibnamefont {Spero}},\ and\ \bibinfo
  {author} {\bibfnamefont {B.}~\bibnamefont {Klipstein}},\ }\bibfield  {title}
  {\bibinfo {title} {Overview of the {LISA} phasemeter},\ }\bibfield  {journal}
  {\bibinfo  {journal} {AIP Conference Proceedings}\ }\textbf {\bibinfo
  {volume} {873}},\ \href {https://doi.org/10.1063/1.2405113}
  {10.1063/1.2405113} (\bibinfo {year} {2006})\BibitemShut {NoStop}%
\bibitem [{\citenamefont {Barke}\ \emph {et~al.}(2014)\citenamefont {Barke},
  \citenamefont {Brause}, \citenamefont {Bykov}, \citenamefont {Esteban},
  \citenamefont {Enggaard}, \citenamefont {Gerberding}, \citenamefont
  {Heinzel}, \citenamefont {Kullmann}, \citenamefont {Pedersen},\ and\
  \citenamefont {Rasmussen}}]{FinalReport}%
  \BibitemOpen
  \bibfield  {author} {\bibinfo {author} {\bibfnamefont {S.}~\bibnamefont
  {Barke}}, \bibinfo {author} {\bibfnamefont {N.}~\bibnamefont {Brause}},
  \bibinfo {author} {\bibfnamefont {I.}~\bibnamefont {Bykov}}, \bibinfo
  {author} {\bibfnamefont {J.~J.}\ \bibnamefont {Esteban}}, \bibinfo {author}
  {\bibfnamefont {A.}~\bibnamefont {Enggaard}}, \bibinfo {author}
  {\bibfnamefont {O.}~\bibnamefont {Gerberding}}, \bibinfo {author}
  {\bibfnamefont {G.}~\bibnamefont {Heinzel}}, \bibinfo {author} {\bibfnamefont
  {J.}~\bibnamefont {Kullmann}}, \bibinfo {author} {\bibfnamefont {S.~M.}\
  \bibnamefont {Pedersen}},\ and\ \bibinfo {author} {\bibfnamefont
  {T.}~\bibnamefont {Rasmussen}},\ }\href@noop {} {\emph {\bibinfo {title}
  {{LISA} Metrology System Final Report}}},\ \bibinfo {type} {Tech. Rep.}\
  \bibinfo {number} {AO/1-6238/10/NL/HB}\ (\bibinfo {year} {2014})\ \bibinfo
  {note}
  {\url{https://pure.mpg.de/pubman/faces/ViewItemFullPage.jsp?itemId=item_2058697_3}}\BibitemShut
  {NoStop}%
\bibitem [{\citenamefont {Gerberding}\ \emph {et~al.}(2015)\citenamefont
  {Gerberding}, \citenamefont {Diekmann}, \citenamefont {Kullmann},
  \citenamefont {Tröbs}, \citenamefont {Bykov}, \citenamefont {Barke},
  \citenamefont {Brause}, \citenamefont {Esteban}, \citenamefont {Schwarze},
  \citenamefont {Reiche} \emph {et~al.}}]{Gerberding2015}%
  \BibitemOpen
  \bibfield  {author} {\bibinfo {author} {\bibfnamefont {O.}~\bibnamefont
  {Gerberding}}, \bibinfo {author} {\bibfnamefont {C.}~\bibnamefont
  {Diekmann}}, \bibinfo {author} {\bibfnamefont {J.}~\bibnamefont {Kullmann}},
  \bibinfo {author} {\bibfnamefont {M.}~\bibnamefont {Tröbs}}, \bibinfo
  {author} {\bibfnamefont {I.}~\bibnamefont {Bykov}}, \bibinfo {author}
  {\bibfnamefont {S.}~\bibnamefont {Barke}}, \bibinfo {author} {\bibfnamefont
  {N.~C.}\ \bibnamefont {Brause}}, \bibinfo {author} {\bibfnamefont {J.~J.}\
  \bibnamefont {Esteban}}, \bibinfo {author} {\bibfnamefont {T.~S.}\
  \bibnamefont {Schwarze}}, \bibinfo {author} {\bibfnamefont {J.}~\bibnamefont
  {Reiche}}, \emph {et~al.},\ }\bibfield  {title} {\bibinfo {title} {Readout
  for intersatellite laser interferometry: Measuring low frequency phase
  fluctuations of high-frequency signals with microradian precision},\
  }\bibfield  {journal} {\bibinfo  {journal} {Review of Scientific
  Instruments}\ }\href {https://doi.org/10.1063/1.4927071} {10.1063/1.4927071}
  (\bibinfo {year} {2015})\BibitemShut {NoStop}%
\bibitem [{\citenamefont {Bayle}\ \emph {et~al.}(2021)\citenamefont {Bayle},
  \citenamefont {Hartwig},\ and\ \citenamefont {Staab}}]{Bayle2021}%
  \BibitemOpen
  \bibfield  {author} {\bibinfo {author} {\bibfnamefont {J.-B.}\ \bibnamefont
  {Bayle}}, \bibinfo {author} {\bibfnamefont {O.}~\bibnamefont {Hartwig}},\
  and\ \bibinfo {author} {\bibfnamefont {M.}~\bibnamefont {Staab}},\ }\bibfield
   {title} {\bibinfo {title} {Adapting time-delay interferometry for {LISA}
  data in frequency},\ }\href {https://doi.org/10.1103/PhysRevD.104.023006}
  {\bibfield  {journal} {\bibinfo  {journal} {Phys. Rev. D}\ }\textbf {\bibinfo
  {volume} {104}},\ \bibinfo {pages} {023006} (\bibinfo {year}
  {2021})}\BibitemShut {NoStop}%
\bibitem [{\citenamefont {Bayle}\ \emph {et~al.}(2019)\citenamefont {Bayle},
  \citenamefont {Lilley}, \citenamefont {Petiteau},\ and\ \citenamefont
  {Halloin}}]{Bayle2019}%
  \BibitemOpen
  \bibfield  {author} {\bibinfo {author} {\bibfnamefont {J.-B.}\ \bibnamefont
  {Bayle}}, \bibinfo {author} {\bibfnamefont {M.}~\bibnamefont {Lilley}},
  \bibinfo {author} {\bibfnamefont {A.}~\bibnamefont {Petiteau}},\ and\
  \bibinfo {author} {\bibfnamefont {H.}~\bibnamefont {Halloin}},\ }\bibfield
  {title} {\bibinfo {title} {Effect of filters on the time-delay interferometry
  residual laser noise for {LISA}},\ }\href
  {https://doi.org/10.1103/PhysRevD.99.084023} {\bibfield  {journal} {\bibinfo
  {journal} {Phys. Rev. D}\ }\textbf {\bibinfo {volume} {99}},\ \bibinfo
  {pages} {084023} (\bibinfo {year} {2019})}\BibitemShut {NoStop}%
\bibitem [{\citenamefont {Hellings}(2001)}]{ClockHellings}%
  \BibitemOpen
  \bibfield  {author} {\bibinfo {author} {\bibfnamefont {R.~W.}\ \bibnamefont
  {Hellings}},\ }\bibfield  {title} {\bibinfo {title} {Elimination of clock
  jitter noise in spaceborne laser interferometers},\ }\href
  {https://doi.org/10.1103/PhysRevD.64.022002} {\bibfield  {journal} {\bibinfo
  {journal} {Phys. Rev. D}\ }\textbf {\bibinfo {volume} {64}},\ \bibinfo
  {pages} {022002} (\bibinfo {year} {2001})}\BibitemShut {NoStop}%
\bibitem [{\citenamefont {Tinto}\ \emph
  {et~al.}(2002{\natexlab{b}})\citenamefont {Tinto}, \citenamefont
  {Estabrook},\ and\ \citenamefont {Armstrong}}]{ClockTinto1}%
  \BibitemOpen
  \bibfield  {author} {\bibinfo {author} {\bibfnamefont {M.}~\bibnamefont
  {Tinto}}, \bibinfo {author} {\bibfnamefont {F.~B.}\ \bibnamefont
  {Estabrook}},\ and\ \bibinfo {author} {\bibfnamefont {J.~W.}\ \bibnamefont
  {Armstrong}},\ }\bibfield  {title} {\bibinfo {title} {Time-delay
  interferometry for lisa},\ }\href
  {https://doi.org/10.1103/PhysRevD.65.082003} {\bibfield  {journal} {\bibinfo
  {journal} {Phys. Rev. D}\ }\textbf {\bibinfo {volume} {65}},\ \bibinfo
  {pages} {082003} (\bibinfo {year} {2002}{\natexlab{b}})}\BibitemShut
  {NoStop}%
\bibitem [{\citenamefont {Tinto}\ and\ \citenamefont
  {Hartwig}(2018)}]{ClockTinto2}%
  \BibitemOpen
  \bibfield  {author} {\bibinfo {author} {\bibfnamefont {M.}~\bibnamefont
  {Tinto}}\ and\ \bibinfo {author} {\bibfnamefont {O.}~\bibnamefont
  {Hartwig}},\ }\bibfield  {title} {\bibinfo {title} {Time-delay interferometry
  and clock-noise calibration},\ }\href
  {https://doi.org/10.1103/PhysRevD.98.042003} {\bibfield  {journal} {\bibinfo
  {journal} {Phys. Rev. D}\ }\textbf {\bibinfo {volume} {98}},\ \bibinfo
  {pages} {042003} (\bibinfo {year} {2018})}\BibitemShut {NoStop}%
\bibitem [{\citenamefont {Hartwig}(2021)}]{HartwigPhD}%
  \BibitemOpen
  \bibfield  {author} {\bibinfo {author} {\bibfnamefont {O.}~\bibnamefont
  {Hartwig}},\ }\emph {\bibinfo {title} {Instrumental Modelling and Noise
  Reduction Algorithms for the Laser Interferometer Space Antenna}},\ \href
  {https://www.repo.uni-hannover.de/handle/123456789/11459} {Ph.D. thesis},\
  \bibinfo  {school} {Leibniz Universität Hannover} (\bibinfo {year}
  {2021})\BibitemShut {NoStop}%
\end{thebibliography}%

\end{document}